\documentclass[aps,prb,showpacs,amsmath,amssymb]{revtex4}
\usepackage{graphicx}
\usepackage{bm}

\def\br{ \bm{r} }
\def\bk{ \bm{k} }
\def\bq{ \bm{q} }

\def\bH{ \bm{H} }
\def\bK{ \bm{K} }
\def\bA{ \bm{A} }
\def\bD{ \bm{D} }
\def\bgam{ \bm{\gamma} }

\def\sign{ \,\mathrm{sign}\, }
\def\tr{ \,\mathrm{tr}\, }
\def\bnab{ \bm{\nabla}}
\def\bsigma{ \hat{\bm{\sigma}} }

\begin{document}

\title{Effects of impurities on $H_{c2}(T)$ in superconductors without
inversion symmetry}

\author{K. V. Samokhin}

\affiliation{Department of Physics, Brock University,
St.Catharines, Ontario L2S 3A1, Canada}
\date{\today}

\begin{abstract}
We calculate the upper critical field, $H_{c2}(T)$, due to the
orbital pair breaking in disordered superconductors without
inversion symmetry. Differences from the usual centrosymmetric
case are highlighted. The linearized gap equations in magnetic
field, with the singlet and triplet pairing channels mixed by
impurity scattering, are solved exactly for a cubic crystal.
\end{abstract}

\pacs{74.20.-z, 74.25.Op}

\maketitle

\section{Introduction}
\label{sec: Intro}

Recently, superconductivity has been discovered in a number of
compounds lacking inversion symmetry, such as CePt$_3$Si (Ref.
\onlinecite{Bauer04}), UIr (Ref. \onlinecite{Akazawa04}),
CeRhSi$_3$ (Ref. \onlinecite{Kimura05}), CeIrSi$_3$ (Ref.
\onlinecite{Sugitani06}), Li$_2$(Pd$_{3-x}$Pt$_{x}$)B (Ref.
\onlinecite{LiPt-PdB}), and many others. Much of the theoretical
work in the field has focussed on searching for the features which
are specific to noncentrosymmetric systems. These include the
magnetoelectric effect,\cite{Lev85,Edel89,Fuji05} a large residual
spin susceptibility and reduced paramagnetic limiting,
\cite{Edel89,GR01,Yip02,FAKS04,Sam07} and various novel nonuniform
superconducting states.\cite{Agter03,Sam04,KAS05,MS08}

In this paper we study the effects of the absence of inversion
symmetry on the upper critical field, $H_{c2}(T)$, at arbitrary
temperature. We assume the pairing to be of the
Bardeen-Cooper-Schrieffer (BCS) type, and include only the orbital
pair breaking. The main qualitative difference from the
centrosymmetric case is that the spin-orbit (SO) coupling of
electrons with the crystal lattice changes the nature of
single-electron states, lifting spin degeneracy of the energy
bands. Then even scalar impurities can mix the singlet and triplet
channels in the Cooper pair propagator, thus making the theory
considerably more complicated. The derivation of the $H_{c2}$
equations for arbitrary noncentrosymmetric crystal symmetry is
presented in Sec. \ref{sec: Hc2 equations} below, with some of the
technical details relegated to Appendices \ref{sec: app A} and
\ref{sec: app B}. In Sec. \ref{sec: cubic}, we apply the general
equations to a cubic superconductor with the point group
$\mathbb{G}=\mathbf{O}$. Assuming that both the band structure and
the SO coupling are fully isotropic, we are able to exactly solve
the coupled equations for the singlet and triplet channels, obtain
the $H_{c2}$ equation in a closed form, and derive analytical
expressions for the upper critical field in the ``dirty'' limit.
This isotropic model clearly shows the deviations from the usual,
i.e. centrosymmetric BCS, case, for which the upper critical field
was calculated in the classic papers by Helfand, Werthamer, and
Hohenberg in 1960s (Refs. \onlinecite{HW66,WHH66}). Sec. \ref{sec:
Conclusions} contains a discussion of our results.

The magnetic phase diagram of noncentrosymmetric superconductors
has been discussed previously in several works. The upper critical
field for a clean three-dimensional Rashba superconductor was
calculated in Ref. \onlinecite{KAS05}, while the effects of
disorder in the Ginzburg-Landau regime were studied in Ref.
\onlinecite{MS07}. Two-dimensional case, in which only the
paramagnetic pair breaking is present, was considered in Ref.
\onlinecite{DF07}. Recently, $H_{c2}$ at all temperatures was
calculated in Ref. \onlinecite{Sam-Hc2}, neglecting the
impurity-induced triplet channel in the pair propagator in the
limit when the SO band splitting is small compared with the Fermi
energy. In this paper, we relax this last condition and include
both the singlet and triplet channels.

Throughout the paper we use the units in which $\hbar=k_B=1$.

\section{Derivation of $H_{c2}$ equations: General case}
\label{sec: Hc2 equations}

Let us consider a noncentrosymmetric superconductor with the
Hamiltonian given by $H=H_0+H_{imp}+H_{int}$. The first term,
\begin{equation}
\label{H_0}
    H_0=\sum\limits_{\bk}[\epsilon_0(\bk)\delta_{\alpha\beta}+
    \bgam(\bk)\bm{\sigma}_{\alpha\beta}]
    a^\dagger_{\bk\alpha}a_{\bk\beta},
\end{equation}
describes non-interacting electrons in the crystal lattice
potential, where $\alpha,\beta=\uparrow,\downarrow$ are spin
indices, $\epsilon_0(\bk)$ is the quasiparticle energy counted
from $\epsilon_F$, and $\hat{\bm{\sigma}}$ are the Pauli matrices.
In Eq. (\ref{H_0}) and everywhere below, summation over repeated
spin indices is implied, while summation over space and band
indices is always shown explicitly. The second term in Eq.
(\ref{H_0}), with $\bgam(\bk)=-\bgam(-\bk)$, describes a
Rashba-type (or antisymmetric) SO coupling of electrons with the
crystal lattice.\cite{Rashba60} In addition, there might be a
usual (symmetric) SO coupling, present even in centrosymmetric
crystals. If the latter is included, then $\alpha,\beta$ in Eq.
(\ref{H_0}) should be interpreted as pseudospin projections.
Diagonalization of $H_0$ yields two non-degenerate bands labelled
by the helicity $\lambda=\pm$:
\begin{equation}
\label{xis gen}
    \xi_\lambda(\bk)=\epsilon_0(\bk)+\lambda|\bgam(\bk)|.
\end{equation}
The Fermi velocities in the two bands are given by
$\bm{v}_\lambda(\bk)=\partial\xi_\lambda/\partial\bk$. The
Fermi-level densities of states are defined in the usual way by
$N_\lambda=\mathcal{V}^{-1}\sum_{\bk}\delta[\xi_\lambda(\bk)]$
(${\cal V}$ is the system volume), and the difference between
$N_+$ and $N_-$ is characterized by a parameter
\begin{equation}
\label{delta def}
    \delta=\frac{N_+-N_-}{N_++N_-}.
\end{equation}
If the SO coupling is small compared with the Fermi energy, then
$\delta\sim\mathcal{O}(E_{SO}/\epsilon_F)$, where
$E_{SO}=2\max_{\bk}|\bgam(\bk)|$ is a measure of the SO band
splitting.

Scattering of electrons at isotropic scalar impurities is
introduced according to
\begin{equation}
\label{H_imp}
    H_{imp}=\int d^3\br\,
    U(\br)\psi^\dagger_\alpha(\br)\psi_\alpha(\br).
\end{equation}
The random potential $U(\br)$ has zero mean and is characterized
by the correlator $\langle
U(\br)U(\br')\rangle=n_{imp}U_0^2\delta(\br-\br')$, where
$n_{imp}$ is the impurity concentration and $U_0$ has the meaning
of the strength of an individual point-like impurity. The field
operators are given by $\psi_\alpha(\br)={\cal
V}^{-1/2}\sum_{\bk}e^{i\bk\br}a_{\bk\alpha}$.

Neglecting the paramagnetic pair breaking, which is a good
assumption in many bulk noncentrosymmetric materials, the effect
of a uniform external magnetic field $\bH$ is described by the
Peierls substitution:\cite{LL9}
\begin{equation}
\label{hat h}
    \hat{h}=\epsilon_0(\bK)+\bgam(\bK)\bsigma+U(\br),
\end{equation}
where $\bK=-i\bm{\nabla}+(e/c)\bA(\br)$, and $e$ is the absolute
value of the electron charge.

We describe the pairing interaction by a BCS-like Hamiltonian:
\begin{equation}
\label{H_int}
    H_{int}=-V\int d^3\br\,\psi_\uparrow^\dagger(\br)\psi_\downarrow^\dagger(\br)
    \psi_\downarrow(\br)\psi_\uparrow(\br),
\end{equation}
where $V>0$ is the coupling constant. In this model, the
superconducting order parameter is represented by a single complex
function $\eta(\br)$, see Ref. \onlinecite{SM08}. The critical
temperature at a given field, or inversely the upper critical
field, $H_{c2}(T)$, at a given temperature, is found from the
condition that the linearized gap equation
\begin{equation}
\label{linear gap eq}
    \biggl[\frac{1}{V}-T\sum_n{}^\prime\hat X(\omega_n)\biggr]\eta(\br)=0
\end{equation}
has a nontrivial solution. Here $\omega_n=(2n+1)\pi T$ is the
fermionic Matsubara frequency, the prime in the second term means that the summation is limited to
$|\omega_n|\leq\omega_c$, where $\omega_c$ is the BCS frequency
cutoff, and the operator $\hat X(\omega_n)$ is defined by the following kernel:
\begin{equation}
\label{X def}
    X(\br,\br';\omega_n)=\frac{1}{2}\bigl\langle\tr\hat g^\dagger\hat G(\br,\br';\omega_n)
    \hat g\hat G^T(\br,\br';-\omega_n)\bigr\rangle_{imp},
\end{equation}
where $\hat g=i\hat\sigma_2$. The angular brackets denote the impurity averaging, and
$\hat G(\br,\br';\omega_n)$ is the Matsubara Green's functions of
electrons in the normal state, which satisfies the equation
\begin{equation}
\label{G eq gen}
    (i\omega_n-\hat{h})\hat
    G(\br,\br';\omega_n)=\delta(\br-\br'),
\end{equation}
where the single-particle Hamiltonian $\hat{h}$ is given by
expression (\ref{hat h}).

At zero field, Eq. (\ref{G eq gen}) yields the following
expression for the average Green's function:
\begin{equation}
\label{G zero H}
    \hat{G}_0(\bk,\omega_n)=
    \sum_{\lambda=\pm}\hat\Pi_\lambda(\bk)G_\lambda(\bk,\omega_n),
\end{equation}
where
\begin{equation}
\label{Pis}
    \hat\Pi_\lambda(\bk)=\frac{1+\lambda\hat\bgam(\bk)\bsigma}{2}
\end{equation}
are the band projection operators ($\hat\bgam=\bgam/|\bgam|$), and
\begin{equation}
\label{G lambda average zero H}
    G_\lambda(\bk,\omega_n)=
    \frac{1}{i\omega_n-\xi_\lambda(\bk)+i\Gamma\sign\omega_n},
\end{equation}
are the electron Green's functions in the band representation. Here
$\xi_\lambda(\bk)$ is the quasiparticle dispersion in the $\lambda$th band, see Eq. (\ref{xis gen}),
$\Gamma=1/2\tau$ is the elastic scattering rate, $\tau=(2\pi
n_{imp}U_0^2N_F)^{-1}$ is the electron mean free time due to
impurities, and
\begin{equation}
\label{NF def}
    N_F=\frac{N_++N_-}{2}.
\end{equation}

\begin{figure}[t]
    \includegraphics[width=8.1cm]{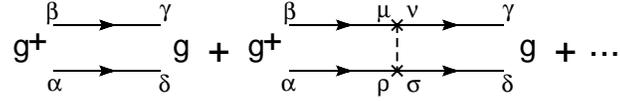}
    \caption{Impurity ladder diagrams in the Cooper channel. Lines with
    arrows correspond to the average Green's functions of electrons,
    $\hat g=i\hat\sigma_2$, and the impurity (dashed) lines
    are defined in the text, see Eq. (\ref{impurity line}).}
    \label{fig: ladder diagrams}
\end{figure}

The impurity average of the product of two Green's functions in
Eq. (\ref{X def}) can be represented graphically by the ladder
diagrams, see Fig. \ref{fig: ladder diagrams}. We assume the
disorder to be sufficiently weak for the diagrams with crossed
impurity lines to be negligible, see Ref. \onlinecite{AGD}. In
order to solve Eq. (\ref{linear gap eq}) at nonzero field, we
introduce an impurity-renormalized gap function $\hat
D(\br,\omega_n)$, which a matrix in the spin space satisfying the
following integral equation:
\begin{eqnarray}
\label{D eq}
    \hat D(\br,\omega_n)=\eta(\br)\hat g+\frac{1}{2}n_{imp}U_0^2\hat g
    \int d^3\br'\tr\hat g^\dagger\hat
    G(\br,\br';\omega_n)\hat D(\br',\omega_n)\hat G^T(\br,\br';-\omega_n)\nonumber\\
    +\frac{1}{2}n_{imp}U_0^2\hat{\bm{g}}\int d^3\br'\tr\hat{\bm{g}}^\dagger
    \hat G(\br,\br';\omega_n)\hat D(\br',\omega_n)\hat G^T(\br,\br';-\omega_n),
\end{eqnarray}
where $\hat G(\br,\br';\omega_n)$ are the disorder-averaged
solutions of Eq. (\ref{G eq gen}). The above equation can be
easily derived from the impurity ladder diagrams in Fig. \ref{fig:
ladder diagrams}, by representing each ``rung'' of the ladder as a
sum of spin-singlet and spin-triplet terms:
\begin{equation}
\label{impurity line}
    n_{imp}U_0^2\delta_{\mu\nu}\delta_{\rho\sigma}=
    \frac{1}{2}n_{imp}U_0^2g_{\mu\rho}g^\dagger_{\sigma\nu}+
    \frac{1}{2}n_{imp}U_0^2\bm{g}_{\mu\rho}\bm{g}^\dagger_{\sigma\nu},
\end{equation}
where $\hat{\bm{g}}=i\hat{\bm{\sigma}}\hat\sigma_2$.

Seeking solution of Eq. (\ref{D eq}) in the form
\begin{equation}
\label{D st}
    \hat D(\br,\omega_n)=d_0(\br,\omega_n)\hat
    g+\bm{d}(\br,\omega_n)\hat{\bm{g}},
\end{equation}
we obtain a system of four integral equations for
$d_a(\br,\omega_n)$, where $a=0,1,2,3$:
\begin{equation}
\label{dd eqs}
    \sum_{b=0}^3\bigl[\delta_{ab}-\Gamma\hat{\mathcal{Y}}_{ab}(\omega_n)\bigr]
    d_b(\br,\omega_n)=\eta(\br)\delta_{a0}.
\end{equation}
Here the operators $\hat{\mathcal{Y}}_{ab}(\omega_n)$ are defined
by the kernels
\begin{equation}
\label{Yab def}
    \mathcal{Y}_{ab}(\br,\br';\omega_n)=\frac{1}{2\pi N_F}\tr\hat{\mathrm{g}}_a^\dagger
    \hat G(\br,\br';\omega_n)\hat{\mathrm{g}}_b\hat G^T(\br,\br';-\omega_n),
\end{equation}
with $\hat{\mathrm{g}}_0=\hat g$, and $\hat{\mathrm{g}}_i=\hat
g_i$ for $i=1,2,3$. We see that, in addition to the spin-singlet
component $d_0(\br,\omega_n)$, impurity scattering can induce also
a nonzero spin-triplet component $\bm{d}(\br,\omega_n)$. The gap
equation (\ref{linear gap eq}) contains only the singlet
component: Using Eqs. (\ref{dd eqs}), we obtain:
\begin{equation}
\label{gap eq d0}
    \frac{1}{N_FV}\eta(\br)-\pi T\sum_n{}'\frac{d_0(\br,\omega_n)-\eta(\br)}{\Gamma}=0.
\end{equation}
We would like to note that the triplet component does not appear
in the centrosymmetric case. Indeed, in the absence of the Zeeman
interaction the spin structure of the Green's function is trivial:
$G_{\alpha\beta}(\br,\br';\omega_n)=\delta_{\alpha\beta}G(\br,\br';\omega_n)$.
Then it follows from Eq. (\ref{Yab def}) that
$\hat{\mathcal{Y}}_{ab}(\omega_n)=\delta_{ab}\hat{\mathcal{Y}}(\omega_n)$,
therefore $d_0=(1-\Gamma\hat{\mathcal{Y}})^{-1}\eta$ and
$\bm{d}=0$.

The next step is to find the spectrum of the operators
$\hat{\mathcal{Y}}_{ab}(\omega_n)$. The orbital effect of the
magnetic field is described by a phase factor in the average
electron Green's function: $\hat G(\br,\br';\omega_n)=\hat
G_0(\br-\br';\omega_n)e^{i\varphi(\br,\br')}$, where $\hat G_0$ is
the average Green's function in the normal state at zero field,
$\varphi(\br,\br')=(e/c)\int_{\br}^{\br'}\bm{A}(\br)d\br$, and the
integration is performed along a straight line connecting $\br$
and $\br'$ (Ref. \onlinecite{AGD}). The ``phase-only''
approximation is legitimate if the temperature is not very low, so
that the Landau level quantization can be neglected. Using the
identity
$e^{2i\varphi(\br,\br')}\eta(\br')=e^{-i(\br-\br')\bD}\eta(\br)$,
where $\bD=-i\bnab+(2e/c)\bm{A}$, we obtain:
\begin{equation}
\label{q to D}
    \hat{\mathcal{Y}}_{ab}(\omega_n)=\bar{\mathcal{Y}}_{ab}(\bq,\omega_n)\bigr|_{\bq\to\bD},
\end{equation}
where
\begin{equation}
\label{Yab}
    \bar{\mathcal{Y}}_{ab}(\bq,\omega_n)=\frac{1}{2\pi N_F}\int\frac{d^3\bk}{(2\pi)^3}\tr
    \hat{\mathrm{g}}_a^\dagger\hat G_0(\bk+\bq,\omega_n)
    \hat{\mathrm{g}}_b\hat G_0^T(-\bk,-\omega_n),
\end{equation}
Substituting here the Green's functions (\ref{G zero H}) and
calculating the spin traces, we obtain for the singlet-singlet
term:
\begin{equation}
\label{Y00}
    \bar{\mathcal{Y}}_{00}(\bq,\omega_n)=\frac{1}{2}\sum_\lambda\rho_\lambda
    \left\langle\frac{1}{|\omega_n|+\Gamma+i\bm{v}_\lambda(\bk)\bq\sign\omega_n/2}
    \right\rangle_\lambda,
\end{equation}
where
\begin{equation}
\label{rhos}
    \rho_\pm=\frac{N_\pm}{N_F}=1\pm\delta
\end{equation}
are the fractional densities of states in the two bands, and
$\langle(...)\rangle_\lambda$ denotes the Fermi-surface averaging
in the $\lambda$th band. Similarly, for the singlet-triplet mixing
terms we obtain:
\begin{equation}
\label{Y0i}
    \bar{\mathcal{Y}}_{0i}(\bq,\omega_n)=\bar{\mathcal{Y}}_{i0}(\bq,\omega_n)
    =\frac{1}{2}\sum_\lambda\lambda\rho_\lambda\left\langle\frac{\hat\gamma_i(\bk)}{|\omega_n|
    +\Gamma+i\bm{v}_\lambda(\bk)\bq\sign\omega_n/2}\right\rangle_\lambda.
\end{equation}
We see that the mixing occurs due to the SO coupling and vanishes
at $\bgam\to 0$, when $\rho_+=\rho_-=1$ and
$\bm{v}_+=\bm{v}_-=\bm{v}_F$. Finally, the triplet-triplet terms
can be represented as follows:
\begin{equation}
\label{Yij}
    \bar{\mathcal{Y}}_{ij}(\bq,\omega_n)=\bar{\mathcal{Y}}^{(1)}_{ij}(\bq,\omega_n)+\bar{\mathcal{Y}}^{(2)}_{ij}(\bq,\omega_n),
\end{equation}
where
\begin{equation}
\label{Y1ij}
    \bar{\mathcal{Y}}^{(1)}_{ij}(\bq,\omega_n)=\frac{1}{2}\sum_\lambda\rho_\lambda\left\langle
    \frac{\hat\gamma_i(\bk)\hat\gamma_j(\bk)}{|\omega_n|+\Gamma+i\bm{v}_\lambda(\bk)\bq\sign\omega_n/2}
    \right\rangle_\lambda,
\end{equation}
and
\begin{equation}
\label{Y2ij}
    \bar{\mathcal{Y}}^{(2)}_{ij}(\bq,\omega_n)=\frac{1}{2\pi N_F}\sum_\lambda\int\frac{d^3\bk}{(2\pi)^3}
    (\delta_{ij}-\hat\gamma_i\hat\gamma_j-i\lambda e_{ijl}\hat\gamma_l)
    G_\lambda(\bk+\bq,\omega_n)G_{-\lambda}(-\bk,-\omega_n).
\end{equation}
The singlet impurity scattering channel, which is described by the
first term in expression (\ref{impurity line}), causes only the
scattering of intraband pairs between the bands. In contrast, the
triplet impurity scattering can create also interband pairs, which
are described by $\bar{\mathcal{Y}}^{(2)}_{ij}$. It is easy to
show that if the SO band splitting exceeds both $\omega_c$ and
$\Gamma$, then the second (interband) term in Eq. (\ref{Yij}) is
smaller than the first (intraband) one, see Appendix \ref{sec: app
A}. Note that in real materials, $E_{SO}$ ranges from tens to
hundreds meV, see Ref. \onlinecite{SZB04} for CePt$_3$Si, and Ref.
\onlinecite{LP05} for Li$_2$Pd$_3$B and Li$_2$Pt$_3$B. On the
other hand, there is still considerable uncertainty as to the
values of $\omega_c$, especially in heavy-fermion compounds, such
as CePt$_3$Si. The typical energy of phonons responsible for the
pairing in Li$_2$Pd$_3$B was estimated in Ref. \onlinecite{LP05}
to be 20 meV, while the SO band splitting is 30 meV (reaching 200
meV in Li$_2$Pt$_3$B).

The critical temperature of the phase transition into a uniform
superconducting state at zero field can be found by setting
$\bq=0$ in the above expressions. According to Eq. (\ref{Y0i}),
the singlet and triplet channels are decoupled. Then it follows
from Eqs. (\ref{Y00}) and (\ref{dd eqs}) that
$d_0(\omega_n)=(1+\Gamma/|\omega_n|)\eta$. Substituting this into
Eq. (\ref{gap eq d0}), we obtain:
\begin{equation}
\label{Tc0 eq}
    \frac{1}{N_FV}-\pi T\sum_n{}'\frac{1}{|\omega_n|}=0,
\end{equation}
which yields the superconducting critical temperature:
\begin{equation}
\label{Tc0}
    T_{c0}=\frac{2e^{\mathbb{C}}}{\pi}\omega_c e^{-1/N_FV},
\end{equation}
where $\mathbb{C}\simeq 0.577$ is Euler's constant. We see that
there is an analog of Anderson's theorem in noncentrosymmetric
superconductors with a BCS-contact pairing interaction: The
zero-field critical temperature is not affected by scalar
disorder.\cite{MS07}

In the presence of magnetic field, neglecting the interband
contributions to the triplet pair propagator, we obtain:
\begin{equation}
\label{Yab intraband}
    \bar{\mathcal{Y}}_{ab}(\bq,\omega_n)=\frac{1}{2}\sum_\lambda\rho_\lambda\left\langle
    \frac{\Lambda_{\lambda,a}(\bk)\Lambda_{\lambda,b}(\bk)}{|\omega_n|+\Gamma
    +i\bm{v}_\lambda(\bk)\bq\sign\omega_n/2}\right\rangle_\lambda,
\end{equation}
where
\begin{equation}
\label{Lambda def}
    \Lambda_{\lambda,a}(\bk)=\left\{\begin{array}{lll}
      1 & , & a=0 \\
      \lambda\hat\gamma_a(\bk) & , & a=1,2,3 \\
    \end{array}\right..
\end{equation}
Next we use in Eq. (\ref{Yab intraband}) the identity
$x^{-1}=\int_0^\infty du\,e^{-xu}$, and make the substitution
$\bq\to\bD$, see Eq. (\ref{q to D}), in the exponent to represent
$\hat{\mathcal{Y}}_{ab}(\omega_n)$ as a differential operator of
infinite order:
\begin{equation}
\label{hat Yab final}
    \hat{\mathcal{Y}}_{ab}(\omega_n)=\frac{1}{2}\int_0^\infty du\;e^{-u(|\omega_n|+\Gamma)}
    \sum_\lambda\rho_\lambda\hat{\mathcal{O}}^{ab}_\lambda,
\end{equation}
where
\begin{equation}
\label{hat O def}
    \hat{\mathcal{O}}^{ab}_\lambda=\left\langle\Lambda_{\lambda,a}(\bk)\Lambda_{\lambda,b}(\bk)
    e^{-iu\bm{v}_\lambda(\bk)\bD\sign\omega_n/2}\right\rangle_\lambda.
\end{equation}

In order to solve Eqs. (\ref{dd eqs}), with the operators
$\hat{\mathcal{Y}}_{ab}(\omega_n)$ given by expressions (\ref{hat
Yab final}), we follow the procedure described in Ref.
\onlinecite{HW66}. We choose the $z$-axis along the external
field, so that $\bH=H\hat z$, and introduce the operators
\begin{equation}
\label{a operators}
    a_\pm=\ell_H\frac{D_x\pm iD_y}{2},\quad
    a_3=\ell_HD_z,
\end{equation}
where $\ell_H=\sqrt{c/eH}$ is the magnetic length. It is easy to
check that $a_+=a_-^\dagger$ and $[a_-,a_+]=1$, therefore $a_\pm$
have the meaning of the raising and lowering operators, while
$a_3=a_3^\dagger$ commutes with both of them: $[a_3,a_\pm]=0$. It
is convenient to expand both the order parameter $\eta$ and the
impurity-renormalized gap functions $d_a$ in the basis of Landau
levels $|N,p\rangle$, which satisfy
\begin{equation}
\label{a Np}
    a_+|N,p\rangle=\sqrt{N+1}|N+1,p\rangle,\qquad
    a_-|N,p\rangle=\sqrt{N}|N-1,p\rangle,\qquad
    a_3|N,p\rangle=p|N,p\rangle,
\end{equation}
where $N=0,1,...$, and $p$ is a real number. We have
\begin{equation}
\label{eta expand}
    \eta(\br)=\sum_{N,p}\eta_{N,p}\langle\br|N,p\rangle,\qquad
    d_a(\br,\omega_n)=\sum_{N,p}d^a_{N,p}(\omega_n)\langle\br|N,p\rangle.
\end{equation}
According to Eqs. (\ref{dd eqs}), the expansion coefficients
satisfy the following algebraic equations:
\begin{equation}
\label{dd eqs linear}
    \sum_{N',p',b}\Bigl[\delta_{ab}\delta_{NN'}\delta_{pp'}
    -\Gamma\langle N,p|\hat{\mathcal{Y}}_{ab}(\omega_n)|N',p'\rangle\Bigr]
    d^b_{N',p'}(\omega_n)=\delta_{a0}\eta_{N,p}.
\end{equation}
Substituting the solutions of these equations into
\begin{equation}
\label{gap eq LL}
    \frac{1}{N_FV}\eta_{N,p}-\pi T\sum_n{}'\frac{d^0_{N,p}(\omega_n)-\eta_{N,p}}{\Gamma}=0,
\end{equation}
see Eq. (\ref{gap eq d0}), and setting the determinant of the
resulting linear equations for $\eta_{N,p}$ to zero, one arrives
at an equation for the upper critical field.

\section{Cubic case}
\label{sec: cubic}

In the general case, i.e. for arbitrary crystal symmetry and
electronic band structure, the procedure outlined in the previous section
does not yield an equation for $H_{c2}(T)$ in a closed form, since all the Landau
levels are coupled, and one has to diagonalize infinite matrices.
In order to make progress, we focus on the case of a
noncentrosymmetric cubic superconductor with the point group
$\mathbb{G}=\mathbf{O}$, which describes, for instance, the
crystal symmetry of Li$_2$(Pd$_{1-x}$,Pt$_x$)$_3$B. The simplest
expression for the SO coupling compatible with all symmetry
requirements has the following form:
\begin{equation}
\label{gamma_O}
    \bgam(\bk)=\gamma_0\bk,
\end{equation}
where $\gamma_0$ is a constant. We assume a parabolic band:
$\epsilon_0(\bk)=\bk^2/2m^*-\epsilon_F$, where $m^*$ is the
effective mass, $\epsilon_F=k_0^2/2m^*$, and $k_0$ is the Fermi
wave vector in the absence of the SO coupling. The band dispersion
functions are given by
\begin{equation}
\label{xis cubic}
    \xi_\lambda(\bk)=\frac{k^2-k_0^2}{2m^*}+\lambda|\gamma_0|k,
\end{equation}
so that the SO band splitting is isotropic and given by
$E_{SO}=2|\gamma_0|k_0$. It is convenient to characterize the SO
coupling strength by a dimensionless parameter
$\varrho=E_{SO}/4\epsilon_F$. While the two Fermi surfaces have
different radii:
$k_{F,\lambda}=k_0(\sqrt{1+\varrho^2}-\lambda\varrho)$, the Fermi
velocities are the same: $\bm{v}_\lambda(\bk)=v_F\hat{\bk}$, where
$v_F=k_0\sqrt{1+\varrho^2}/m^*$. For the parameter $\delta$, which
characterizes the difference between the band densities of states,
see Eq. (\ref{delta def}), we have
$|\delta|=2\varrho\sqrt{1+\varrho^2}/(1+2\varrho^2)$. We assume
that
\begin{equation}
\label{delta range}
    \delta_c\ll|\delta|\leq 1,
\end{equation}
where $\delta_c=\max(\omega_c,\Gamma)/\epsilon_F\ll 1$. While the
first inequality is equivalent to the condition
$E_{SO}\gg\max(\omega_c,\Gamma)$, which ensures the smallness of
the interband contribution to the Cooper impurity ladder (see
Appendix \ref{sec: app A}), the second one is always satisfied,
with $|\delta|\to 1$ corresponding to the rather unrealistic limit
of extremely strong SO coupling, $\varrho\to\infty$.

In order to solve the gap equations, we make a change of variables
in the triplet component:
$$
    d_\pm(\br,\omega_n)=\frac{d_1(\br,\omega_n)\pm id_2(\br,\omega_n)}{\sqrt{2}}.
$$
Then, Eqs. (\ref{dd eqs}) take the following form:
\begin{equation}
\label{dd eqs cubic}
    \left( \begin{array}{cccc}
    1-\Gamma\hat{\mathcal{Y}}_{00} & -\Gamma\hat{\mathcal{Y}}_{03} & -\Gamma\hat{\mathcal{Y}}_{0-} & -\Gamma\hat{\mathcal{Y}}_{0+} \\
    -\Gamma\hat{\mathcal{Y}}_{03} & 1-\Gamma\hat{\mathcal{Y}}_{33} & -\Gamma\hat{\mathcal{Y}}_{3-} & -\Gamma\hat{\mathcal{Y}}_{3+} \\
    -\Gamma\hat{\mathcal{Y}}_{0+} & -\Gamma\hat{\mathcal{Y}}_{3+} & 1-\Gamma\hat{\mathcal{Z}} & -\Gamma\hat{\mathcal{Z}}_{+} \\
    -\Gamma\hat{\mathcal{Y}}_{0-} & -\Gamma\hat{\mathcal{Y}}_{3-} & -\Gamma\hat{\mathcal{Z}}_{-} & 1-\Gamma\hat{\mathcal{Z}}
    \end{array} \right)
    \left( \begin{array}{c}
    d_0 \\ d_3 \\ d_+ \\ d_-
    \end{array} \right)
    =\left( \begin{array}{c}
    \eta \\ 0 \\ 0 \\ 0
    \end{array} \right),
\end{equation}
where
$$
    \hat{\mathcal{Y}}_{0\pm}=\frac{\hat{\mathcal{Y}}_{01}\pm i\hat{\mathcal{Y}}_{02}}{\sqrt{2}},\qquad
    \hat{\mathcal{Y}}_{3\pm}=\frac{\hat{\mathcal{Y}}_{13}\pm i\hat{\mathcal{Y}}_{23}}{\sqrt{2}},\qquad
    \hat{\mathcal{Z}}=\frac{\hat{\mathcal{Y}}_{11}+\hat{\mathcal{Y}}_{22}}{2},\qquad
    \hat{\mathcal{Z}}_{\pm}=\frac{\hat{\mathcal{Y}}_{11}\pm 2i\hat{\mathcal{Y}}_{12}-\hat{\mathcal{Y}}_{22}}{2},
$$
with $\hat{\mathcal{Y}}_{ab}=\hat{\mathcal{Y}}_{ba}$ given by Eqs.
(\ref{hat Yab final}).

According to Sec. \ref{sec: Hc2 equations}, one has to know the
matrix elements of the operators
$\hat{\mathcal{Y}}_{ab}(\omega_n)$ in the basis of the Landau
levels $|N,p\rangle$. After some straightforward algebra, see
Appendix \ref{sec: app B}, we obtain the following expressions for
the nonzero matrix elements:
\begin{eqnarray*}
    &&\langle N,p|\hat{\mathcal{Y}}_{00}(\omega_n)|N,p\rangle=y^{00}_{N,p}(\omega_n),\\
    &&\langle N,p|\hat{\mathcal{Y}}_{03}(\omega_n)|N,p\rangle=y^{03}_{N,p}(\omega_n),\\
    &&\langle N,p|\hat{\mathcal{Y}}_{33}(\omega_n)|N,p\rangle=y^{33}_{N,p}(\omega_n),\\
    &&\langle N,p|\hat{\mathcal{Z}}(\omega_n)|N,p\rangle=z_{N,p}(\omega_n),
\end{eqnarray*}
where
\begin{eqnarray}
\label{y00}
    &&y^{00}_{N,p}(\omega_n)=\int_0^\infty du\,e^{-u(|\omega_n|+\Gamma)}
    \int_0^1 ds\,\cos(pvs)e^{-v^2(1-s^2)/2}L_N[v^2(1-s^2)],\\
\label{y03}
    &&y^{03}_{N,p}(\omega_n)=-i\delta\int_0^\infty du\,e^{-u(|\omega_n|+\Gamma)}
    \int_0^1 ds\,s\sin(pvs)e^{-v^2(1-s^2)/2}L_N[v^2(1-s^2)],\\
\label{y33}
    &&y^{33}_{N,p}(\omega_n)=\int_0^\infty du\,e^{-u(|\omega_n|+\Gamma)}
    \int_0^1 ds\,s^2\cos(pvs)e^{-v^2(1-s^2)/2}L_N[v^2(1-s^2)],\\
\label{z}
    &&z_{N,p}(\omega_n)=\frac{1}{2}\int_0^\infty du\,e^{-u(|\omega_n|+\Gamma)}
    \int_0^1 ds\,(1-s^2)\cos(pvs)e^{-v^2(1-s^2)/2}L_N[v^2(1-s^2)],
\end{eqnarray}
$v=(v_F\sign\omega_n/2\ell_H)u$, and $L_N(x)$ are the Laguerre
polynomials of degree $N$. Similarly, we obtain:
\begin{eqnarray*}
    &&\langle N,p|\hat{\mathcal{Y}}_{0-}(\omega_n)|N+1,p\rangle=
    \langle N+1,p|\hat{\mathcal{Y}}_{0+}(\omega_n)|N,p\rangle=\tilde y^0_{N,p}(\omega_n),\\
    &&\langle N,p|\hat{\mathcal{Y}}_{3-}(\omega_n)|N+1,p\rangle=
    \langle N+1,p|\hat{\mathcal{Y}}_{3+}(\omega_n)|N,p\rangle=\tilde y^3_{N,p}(\omega_n),\\
    &&\langle N,p|\hat{\mathcal{Z}}_-(\omega_n)|N+2,p\rangle=\langle N+2,p|\hat{\mathcal{Z}}_+(\omega_n)|N,p\rangle
    =\tilde z_{N,p}(\omega_n),
\end{eqnarray*}
where
\begin{eqnarray}
\label{tilde y0}
    &&\tilde y^0_{N,p}(\omega_n)=-i\delta\frac{1}{\sqrt{2(N+1)}}\int_0^\infty du\,e^{-u(|\omega_n|+\Gamma)}
    \int_0^1 ds\,v(1-s^2)\cos(pvs)e^{-v^2(1-s^2)/2}L^{(1)}_N[v^2(1-s^2)],\\
\label{tilde y3}
    &&\tilde y^3_{N,p}(\omega_n)=-\frac{1}{\sqrt{2(N+1)}}\int_0^\infty du\,e^{-u(|\omega_n|+\Gamma)}
    \int_0^1 ds\,vs(1-s^2)\sin(pvs)e^{-v^2(1-s^2)/2}L^{(1)}_N[v^2(1-s^2)],\\
\label{tilde z}
    &&\tilde z_{N,p}(\omega_n)=-\frac{1}{2\sqrt{(N+1)(N+2)}}\int_0^\infty du\,e^{-u(|\omega_n|+\Gamma)}
    \int_0^1 ds\,v^2(1-s^2)^2\cos(pvs)e^{-v^2(1-s^2)/2}L^{(2)}_N[v^2(1-s^2)].
\end{eqnarray}
and $L^{(\alpha)}_N(x)$ are the generalized Laguerre polynomials.

It follows from the above expressions that the Landau levels are
decoupled, and for $\eta(\br)=\eta\langle\br|N,p\rangle$ ($\eta$
is a constant) the solution of Eqs. (\ref{dd eqs cubic}) has the
following form:
\begin{equation}
\label{d solution cubic}
    \left(\begin{array}{c}
    d_0(\br,\omega_n)\\
    d_3(\br,\omega_n)\\
    d_+(\br,\omega_n)\\
    d_-(\br,\omega_n)
    \end{array} \right)
    =\left(\begin{array}{c}
    d^0_{N,p}(\omega_n)\langle\br|N,p\rangle\\
    d^3_{N,p}(\omega_n)\langle\br|N,p\rangle\\
    d^+_{N,p}(\omega_n)\langle\br|N+1,p\rangle\\
    d^-_{N,p}(\omega_n)\langle\br|N-1,p\rangle
    \end{array} \right).
\end{equation}
For given $N$ and $p$, the coefficients are found from the equations
\begin{equation}
\label{dNp eqs}
    \sum_{b=0,3,\pm}\mathcal{M}_{ab}(N,p;\omega_n)d^b_{N,p}(\omega_n)=\delta_{a0}\eta,
\end{equation}
where
\begin{equation}
\label{cal M def}
    \hat{\mathcal{M}}(N,p;\omega_n)=\left( \begin{array}{cccc}
    1-\Gamma y^{00}_{N,p} & -\Gamma y^{03}_{N,p} & -\Gamma\tilde y^0_{N,p} & -\Gamma\tilde y^0_{N-1,p} \\
    -\Gamma y^{03}_{N,p} & 1-\Gamma y^{33}_{N,p} & -\Gamma\tilde y^3_{N,p} & -\Gamma\tilde y^3_{N-1,p} \\
    -\Gamma\tilde y^0_{N,p} & -\Gamma\tilde y^3_{N,p} & 1-\Gamma z_{N+1,p} & -\Gamma\tilde z_{N-1,p} \\
    -\Gamma\tilde y^0_{N-1,p} & -\Gamma\tilde y^3_{N-1,p} & -\Gamma\tilde z_{N-1,p} & 1-\Gamma z_{N-1,p}
    \end{array} \right).
\end{equation}
Substituting the solution of Eq. (\ref{dNp eqs}) in Eq. (\ref{gap
eq LL}), and using Eqs. (\ref{Tc0 eq}) and (\ref{Tc0}) to
eliminate both the frequency cutoff and the coupling constant, we
obtain an equation implicitly relating the magnetic field and the
transition temperature at given $N$ and $p$:
\begin{equation}
\label{Hc2 eq final}
    \ln\frac{T_{c0}}{T}=\pi T\sum_n\left\{\frac{1}{|\omega_n|}
    -\frac{[\hat{\mathcal{M}}^{-1}(N,p;\omega_n)]_{00}-1}{\Gamma}\right\}.
\end{equation}
The upper critical field, $H_{c2}(T)$, is obtained by maximizing
the solution of this equation with respect to both $N$ and $p$.

Note that the matrix elements of $\hat{\mathcal{M}}$ which are
responsible for the singlet-triplet mixing, i.e. $y_{N,p}^{03}$,
$\tilde y^0_{N,p}$, and $\tilde y^0_{N-1,p}$, are all proportional
to $\delta$, see Eqs. (\ref{y03}) and (\ref{tilde y0}). Therefore,
at $|\delta|\ll 1$ the singlet and triplet channels are
effectively decoupled. Neglecting the corrections of the order of
$\delta^2$, we obtain from Eq. (\ref{dNp eqs}) that
$[\hat{\mathcal{M}}^{-1}(N,p;\omega_n)]_{00}=(1-\Gamma
y^{00}_{N,p})^{-1}$. Substituting this into Eq. (\ref{Hc2 eq
final}), we recover the Helfand-Werthamer expressions,\cite{HW66}
with the maximum critical field corresponding to $N=p=0$ at all
temperatures. Thus, in the weak SO coupling limit the absence of
inversion symmetry does not bring about any new features in
$H_{c2}(T)$, compared with the centrosymmetric case (as long as
the paramagnetic pair breaking is not included, see Ref.
\onlinecite{Sam-Hc2}).

\subsection{``Dirty'' limit at $N=0$, $p=0$}
\label{sec: N0p0}

At arbitrary magnitude of the SO band splitting, the
singlet-triplet mixing makes the $H_{c2}$ equation in
noncentrosymmetric superconductors considerably more cumbersome
than in the Helfand-Werthamer problem, even in our ``minimal''
isotropic model. It is even possible that, at some values of the
parameters, the maximum critical field is achieved for $N>0$ and
$p\neq 0$, the latter corresponding to a disorder-induced
modulation of the order parameter along the applied field. Leaving
investigation of these exotic possibilities to future work, here
we just consider the case $N=p=0$. Then it follows from Eqs.
(\ref{d solution cubic}) and (\ref{dNp eqs}) that
$d^3_{0,0}=d^-_{0,0}=0$, and
$[\hat{\mathcal{M}}^{-1}(0,0;\omega_n)]_{00}=(1-\Gamma
z_{1,0})/[(1-\Gamma y^{00}_{0,0})(1-\Gamma z_{1,0})
-\Gamma^2(\tilde y^0_{0,0})^2]$. It is convenient to introduce the
reduced temperature, magnetic field, and disorder:
$$
    t=\frac{T}{T_{c0}},\qquad h=\frac{2H}{H_0},\qquad
    \zeta=\frac{\Gamma}{\pi T_{c0}},
$$
where $H_0=\Phi_0/\pi\xi_0^2$, $\Phi_0=\pi c/e$ is the magnetic
flux quantum, and $\xi_0=v_F/2\pi T_{c0}$ is the superconducting
coherence length. In these notations, Eq. (\ref{Hc2 eq final}) yields the following equation
for the upper critical field $h_{c2}(t)$:
\begin{equation}
\label{hc2 eq reduced}
    \ln\frac{1}{t}=2\sum_{n\geq 0}\left[\frac{1}{2n+1}-t
    \frac{w_n(1-\zeta p_n)-\zeta\delta^2q_n^2}{(1-\zeta w_n)(1-\zeta p_n)+\zeta^2\delta^2q_n^2}\right],
\end{equation}
where
\begin{eqnarray}
\label{w p q}
        &&w_n(t,h)=\int_0^\infty d\rho\,e^{-\tilde\omega_n\rho}\int_0^1 ds\,
    e^{-h\rho^2(1-s^2)/4},\nonumber\\
    &&p_n(t,h)=\int_0^\infty d\rho\,e^{-\tilde\omega_n\rho}\int_0^1 ds\,
    \frac{1-s^2}{2}\left[1-\frac{h}{2}\rho^2(1-s^2)\right]e^{-h\rho^2(1-s^2)/4},\\
    &&q_n(t,h)=\int_0^\infty d\rho\,e^{-\tilde\omega_n\rho}\int_0^1 ds\,
    \sqrt{\frac{h}{4}}\rho(1-s^2)e^{-h\rho^2(1-s^2)/4},\nonumber
\end{eqnarray}
where $\tilde\omega_n=(2n+1)t+\zeta$.

In the clean limit, i.e. at $\zeta\to 0$, or if the SO band
splitting is negligibly small, i.e. at $\delta\to 0$, one recovers
from Eq. (\ref{hc2 eq reduced}) the Helfand-Werthamer equation for
a centrosymmetric superconductor. Thus the absence of inversion
symmetry affects the upper critical field only if disorder is
present. One can expect that the effect will be most pronounced in
the ``dirty'' limit, $\zeta\gg 1$. [Note that, according to Eq.
(\ref{delta range}), the disorder strength should satisfy
$\zeta\ll(\epsilon_F/T_{c0})|\delta|$.] We shall see that in this
limit $h_{c2}$ scales as $\zeta$, which allows one to use the
Taylor expansions of the exponentials in Eqs. (\ref{w p q}):
$$
    w_n(t,h)\simeq
    \frac{1}{\tilde\omega_n}\left(1-\frac{h}{3\tilde\omega_n^2}\right),\qquad
    p_n(t,h)\simeq
    \frac{1}{3\tilde\omega_n}\left(1-\frac{6h}{5\tilde\omega_n^2}\right),\qquad
    q_n(t,h)\simeq \frac{\sqrt{h}}{3\tilde\omega_n^2}.
$$
Using the fact that the main contribution to the Matsubara sum in
Eq. (\ref{hc2 eq reduced}) comes from $(2n+1)t\ll\zeta$, we arrive
at a well-known universal equation, which describes the magnetic
pair breaking in superconductors:\cite{Tink-book}
\begin{equation}
\label{universal eq}
    \ln\frac{1}{t}=\Psi\left(\frac{1}{2}+\frac{\sigma}{t}\right)
    -\Psi\left(\frac{1}{2}\right),
\end{equation}
where $\Psi(x)$ is the digamma function, and
\begin{equation}
\label{pair breaking}
    \sigma=\frac{2+\delta^2}{12\zeta}h
\end{equation}
characterizes the pair-breaker strength. Note that the
corresponding expression in the centrosymmetric case is different:
$\sigma_{CS}=h/6\zeta$ (Ref. \onlinecite{HW66}). Analytical
expressions for the upper critical field can be obtained in the
weak-field limit near the critical temperature:
\begin{equation}
\label{hc2 dirty low h}
    h_{c2}|_{t\to 1}=\frac{24\zeta}{(2+\delta^2)\pi^2}(1-t),
\end{equation}
and also at low temperatures:
\begin{equation}
\label{hc2 dirty low t}
    h_{c2}|_{t=0}=\frac{3e^{\mathbb{-C}}}{2+\delta^2}\zeta.
\end{equation}
We see that the SO band splitting in the noncentrosymmetric case enhances the orbital pair breaking.

\section{Conclusions}
\label{sec: Conclusions}

We have derived equations for the upper critical field in
noncentrosymmetric superconductors, assuming a BCS-contact pairing
interaction and orbital pair breaking. In a cubic crystal (the
point group $\mathbb{G}=\mathbf{O}$), in which both the electron
dispersion and the SO coupling are isotropic, the gap equations
are shown to be diagonal in the Landau level basis, with the
singlet and triplet channels in the pair propagator mixed
together. For the order parameter corresponding to the lowest
Landau level, without any modulation along the applied field, we
have derived the $H_{c2}$ equation in a closed form and solved it
in the ``dirty'' limit, in which the effects of the absence of
inversion symmetry are expected to be most pronounced. The effect
on the upper critical field of the singlet-triplet mixing, which
is responsible for the deviations from the Helfand-Werthamer
theory, is found to be proportional to $\delta^2$.

Application of our theory to real noncentrosymmetric
superconductors of cubic symmetry, such as
Li$_2$(Pd$_{3-x}$Pt$_{x}$)B, is complicated by the fact that the
Fermi surfaces as well as the SO band splitting are strongly
anisotropic. Using the maximum values of the SO band splitting
from Ref. \onlinecite{LP05}, one can estimate the corrections to
$H_{c2}(T)$ due to the singlet-triplet mixing to be of the order
of several percent.

\section*{Acknowledgments}

This work was supported by a Discovery Grant from the Natural Sciences and Engineering Research Council of Canada.

\appendix

\section{Interband vs intraband contributions}
\label{sec: app A}

In this appendix, we estimate the relative magnitudes of the
intraband and interband contributions to the triplet pair
propagator, Eq. (\ref{Yij}), in the limit when the SO coupling is
strong compared with both the cutoff energy $\omega_c$ and the
elastic scattering rate $\Gamma$. Let us consider an isotropic
band with $\bgam(\bk)=\gamma_0\bk$ in a cubic crystal. Neglecting
for simplicity the differences between the densities of states and
the Fermi velocities in the two bands: $\rho_+=\rho_-=1$ and
$\bm{v}_+=\bm{v}_-=\bm{v}_F$, and setting $\bq=0$, we obtain from
Eqs. (\ref{Y1ij}) and (\ref{Y2ij}):
$$
    \bar{\mathcal{Y}}^{(1)}_{ij}(\bq=0,\omega_n)=\frac{\left\langle\hat\gamma_i\hat\gamma_j\right\rangle_{\hat{\bk}}}{|\omega_n|+\Gamma}
    =\frac{\delta_{ij}}{3(|\omega_n|+\Gamma)}\equiv\mathcal{Y}_{intra}(\omega_n)\delta_{ij},
$$
and
$$
    \bar{\mathcal{Y}}^{(2)}_{ij}(\bq=0,\omega_n)=\frac{1}{2}\sum_\lambda\left\langle
    \frac{\delta_{ij}-\hat\gamma_i\hat\gamma_j}{|\omega_n|+\Gamma
    +i\lambda|\bgam|\sign\omega_n}\right\rangle_{\hat{\bk}}=
    \frac{2\delta_{ij}}{3(|\omega_n|+\Gamma)(1+r^2)}\equiv\mathcal{Y}_{inter}(\omega_n)\delta_{ij},
$$
where $r(\omega_n)=E_{SO}/2(|\omega_n|+\Gamma)$. Due to the BCS
cutoff, the maximum value of $\omega_n$ is equal to $\omega_c$,
therefore $r_{min}\sim E_{SO}/\max(\omega_c,\Gamma)\gg 1$. From
this it follows that
$$
    \max_n\frac{\mathcal{Y}_{inter}(\omega_n)}{\mathcal{Y}_{intra}(\omega_n)}=\frac{2}{1+r_{min}^2}\sim
    \left[\frac{\max(\omega_c,\Gamma)}{E_{SO}}\right]^2\ll 1.
$$
Therefore the interband contribution is small compared with the
intraband one, at all Matsubara frequencies.

\section{Calculation of $\langle N,p|\hat{\mathcal{Y}}_{ab}(\omega_n)|N',p'\rangle$}
\label{sec: app B}

The operators $\hat{\mathcal{Y}}_{ab}(\omega_n)$ are given by
expressions (\ref{hat Yab final}). For a spherical Fermi surface
and $\bgam(\bk)=\gamma_0\bk$, we obtain from Eq. (\ref{hat O
def}):
\begin{equation}
    \hat{\cal O}^{ab}_\lambda=\frac{1}{2}\int_0^\pi
    d\theta\sin\theta\,e^{-iva_3\cos\theta}\int_0^{2\pi}\frac{d\phi}{2\pi}\,
    \Phi^{ab}_\lambda(\theta,\phi)e^{-iv(e^{-i\phi}a_++e^{i\phi}a_-)\sin\theta},
\end{equation}
where $v=(v_F\sign\omega_n/2\ell_H)u$, and
$\Phi^{ab}_\lambda(\theta,\phi)=\Lambda_{\lambda,a}(\bk)\Lambda_{\lambda,b}(\bk)$,
with $\Lambda_{\lambda,0}(\bk)=1$ and
$\Lambda_{\lambda,i}(\bk)=\lambda\hat k_i$ for $i=1,2,3$, see Eq.
(\ref{Lambda def}). Using the well-known operator identity
$e^{A+B}=e^{-[A,B]/2}e^Ae^B$, which holds if the commutator of $A$
and $B$ is a $c$-number, and expanding the exponentials in powers
of $a_{\pm}$, we obtain:
\begin{equation}
\label{Y ab Np}
    \hat{\mathcal{Y}}_{ab}(\omega_n)=\frac{1}{4}\sum_\lambda\rho_\lambda\int_0^\infty du\;e^{-u(|\omega_n|+\Gamma)}
        \int_0^\pi d\theta\sin\theta\,e^{-iva_3\cos\theta}e^{-(v^2/2)\sin^2\theta}
    \hat{\cal L}^{ab}_\lambda(\theta),
\end{equation}
where
\begin{equation}
\label{cal L ab}
    \hat{\cal L}^{ab}_\lambda(\theta)=\sum_{n,m=0}^\infty\frac{(-iv\sin\theta)^{n+m}}{n!m!}
    \left[\int_0^{2\pi}\frac{d\phi}{2\pi}\,\Phi^{ab}_\lambda(\theta,\phi)e^{i(m-n)\phi}\right]a_+^na_-^m.
\end{equation}
Below we perform the detailed calculations for
$\hat{\mathcal{Y}}_{00}$ and
$\hat{\mathcal{Y}}_{0-}=(\hat{\mathcal{Y}}_{01}-i\hat{\mathcal{Y}}_{02})/\sqrt{2}$.
Other matrix elements can be considered in a similar
fashion.

\underline{$\hat{\mathcal{Y}}_{00}$}: Since
$\Phi^{00}_\lambda(\theta,\phi)=1$, the $\phi$-integral on the
right-hand side of Eq. (\ref{cal L ab}) is equal to $\delta_{nm}$,
and
$$
    \hat{\cal L}^{00}_\lambda(\theta)=\sum_{n=0}^\infty\frac{(-v^2\sin^2\theta)^n}{(n!)^2}
    a_+^na_-^n.
$$
It is easy to show, using Eqs. (\ref{a Np}), that
$a_+^na_-^n|N,p\rangle=[N!/(N-n)!]|N,p\rangle$ for $n\leq N$, and
zero otherwise. Therefore,
$$
    \hat{\cal L}^{00}_\lambda(\theta)|N,p\rangle=\sum_{n=0}^N\frac{N!}{(n!)^2(N-n)!}(-v^2\sin^2\theta)^n|N,p\rangle=
    L_N(v^2\sin^2\theta)|N,p\rangle,
$$
where $L_N(x)$ is the Laguerre polynomial of degree $N$.
Substituting this into Eq. (\ref{Y ab Np}), using the fact that
$\rho_++\rho_-=2$, and introducing $s=\cos\theta$, we obtain:
$\langle
N,p|\hat{\mathcal{Y}}_{00}(\omega_n)|N,p\rangle=y^{00}_{N,p}(\omega_n)$,
where $y^{00}_{N,p}(\omega_n)$ is given by Eq. (\ref{y00}).

\underline{$\hat{\mathcal{Y}}_{0-}$}: Since
$\Phi^{0-}_\lambda(\theta,\phi)=\lambda\sin\theta
e^{-i\phi}/\sqrt{2}$, one has $m=n+1$ on the right-hand side of
Eq. (\ref{cal L ab}), and
$$
    \hat{\cal L}^{0-}_\lambda(\theta)=\lambda\frac{\sin\theta}{\sqrt{2}}
     \sum_{n=0}^\infty\frac{(-iv\sin\theta)^{2n+1}}{n!(n+1)!}a_+^na_-^{n+1}.
$$
Using
$a_+^na_-^{n+1}|N+1,p\rangle=\sqrt{N+1}[N!/(N-n)!]|N,p\rangle$ (if
$n\leq N$, zero otherwise), we obtain:
\begin{eqnarray*}
    \hat{\cal L}^{0-}_\lambda(\theta)|N+1,p\rangle=-i\lambda\frac{v\sin^2\theta}{\sqrt{2}}
    \sqrt{N+1}\sum_{n=0}^{N}\frac{N!}{(N-n)!}\frac{(-v^2\sin^2\theta)^n}{n!(n+1)!}|N,p\rangle\\
    =-i\lambda\frac{1}{\sqrt{2(N+1)}}v\sin^2\theta\,L^{(1)}_N(v^2\sin^2\theta)|N,p\rangle,
\end{eqnarray*}
where $L^{(\alpha)}_N(x)$ are the generalized Laguerre polynomials:
$$
    L^{(\alpha)}_N(x)=\sum_{n=0}^N\frac{(N+\alpha)!}{(N-n)!}\frac{(-x)^n}{n!(n+\alpha)!},
$$
see Ref. \onlinecite{AS64} (the Laguerre polynomials are recovered
by setting $\alpha=0$). Substituting this into Eq. (\ref{Y ab Np})
and using $\rho_+-\rho_-=2\delta$, we obtain: $\langle
N,p|\hat{\mathcal{Y}}_{0-}(\omega_n)|N+1,p\rangle=\tilde
y^0_{N,p}(\omega_n)$, where $\tilde y^0_{N,p}(\omega_n)$ is given
by Eq. (\ref{tilde y0}).

\end{document}